\documentstyle[aps,prl,twocolumn,epsf]{revtex}

\begin{document}
\draft

\twocolumn[\hsize\textwidth\columnwidth\hsize\csname@twocolumnfalse\endcsname
\title{Elementary Excitations in Quantum Antiferromagnetic Chains: Dyons, Spinons
and Breathers.}
\author{T. Asano$^{a}$, H. Nojiri$^{b}$, Y. Inagaki$^{a}$, Y. Ajiro$^{a}$, L.P.
Regnault$^{c}$ and \underline{J.P. Boucher}$^{d}$.}
\address{$^{a}$Department of Physics, Kyushu University, Fukuoka 812-8581,
Japan;\\ $^{b}$Department of Physics, Okayama University, Okayama
700-8530, Japan; \\$^{c}$DRFMC-SPSMS, Laboratoire de
Magn\'{e}tisme et de Diffraction Neutronique, CEA, \\ F-38054
Grenoble cedex 9, France;\\ $^{d}$Laboratoire de Spectrom\'{e}trie
Physique, Universit\'{e} J. Fourier Grenoble I, BP 87, \\F-38402
Saint Martin d'H\`{e}res cedex, France.}
\date{ November 2001}
\maketitle
\begin{abstract} Considering experimental results
obtained on three prototype compounds, TMMC, CsCoCl$_{3}$ (or
CsCoBr$_{3}$) and Cu Benzoate, we discuss the importance of
non-linear excitations in the physics of quantum (and classical)
antiferromagnetic spin chains.
\end{abstract}


] \narrowtext
\ \
\section{\ Introduction:}

In the last twenty years, various experimental and theoretical
studies have been devoted to antiferromagnetic (AF) chains. One
fascinating problem concerns the nature of the elementary
excitations in such systems. As a crucial ingredient, one has to
refer to the concept of non-linear excitations (NLE). Typical
examples of NLE are provided by the {\it solitons} and the {\it
breathers}. These particular excitations are solutions of the {\it
sine-Gordon} equation, which, in the wide field of the NLE, is to
be considered as a prototype model \cite{Scott}. The first
evidence of fluctuations associated with solitons in AF chains was
obtained on a classical spin system (in the 80's)
\cite{Boucher80}. Recently, evidence for breathers was
experimentally obtained on a quantum spin system \cite{Asano}. A
short review is presented, along which, the role of the NLE in AF
chains is discussed. Classical as well as quantum spin systems are
analyzed. The case of anisotropic spin models is considered and a
relation with the isotropic case is proposed. Quantization effects
are also shown to complete the descriptions of the NLE in AF
chains: for the solitons, this leads to another NLE concept, the
{\it dyons}; for the breathers to a discretization of the
excitation spectrum. Along this paper, we shall refer explicitly
to three compounds: (CH$_{3}$)$_{4}$NMnCl$_{3}$, alias TMMC, two
Co compounds,
CsCoCl$_{3}$ and CsCoBr$_{3}$\cite{Boucher}, and Cu(C$_{6}$H$_{5}$COO)$_{2}$%
.3H$_{2}$O, alias Cu Benzoate \cite{Asano}. All of them are known for very
long to be good models of 1-dimensional antiferromagnetic chains. In TMMC,
the spin value is large, $s=5/2$, and a semi-classical spin description is a
good approach. The other compounds, however, are representative of quantum
spin systems, with $s=1/2$.

\section{TMMC:}

TMMC is known to provide examples of quasi-isotropic Heisenberg AF chains.
At $T_{co}\simeq 20$ K, however, because of the small dipolar anisotropy $D$%
, a crossover occurs, which changes the spin system from isotropic to
easy-plane behaviors \cite{Boucher79}. The spins can be viewed as being
repelled in the XY plane perpendicular to the chain Z direction. Then, if
one applies a magnetic field $H$ in the XY plane - along the Y axis, for
instance - another crossover occurs because of the preferred spin-flop
configuration. In this later phase, the spins are mainly aligned along the X
axis: they form an Ising-like AF state where the soliton regime takes place.
The soliton phase diagram for TMMC is shown in Fig. 1 \cite{Boucher}.

\begin{figure}[tbp]
\centerline{\epsfxsize=70 mm \epsfbox{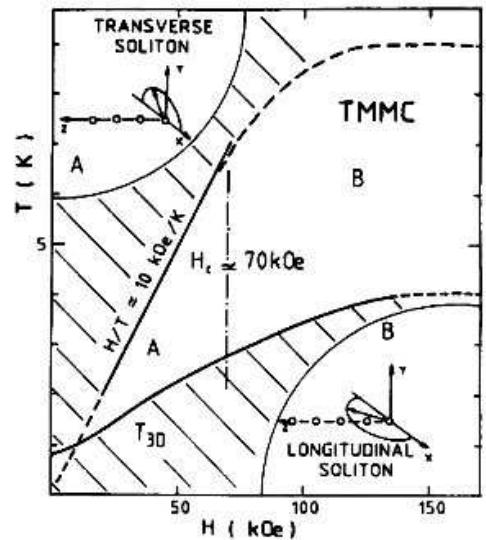}} \vspace{0.5 cm}
\caption{Soliton phase diagram of TMMC [4]. $H_{c}$ is the
cross-over field between the soliton regimes A and B (see the
insets).}
\end{figure}

Two soliton regimes (A and B in Fig. 1, see also the insets) can be defined.
Both they correspond to the sine-Gordon model. The difference between A and
B results from the relative ``competition'' between the two effective
anisotropies, one due to $D$, the other induced by $H$ \cite{Boucher}. We
mainly limit the discussion to the ``transverse'' soliton regime A. For
TMMC, the initial Hamiltonian is given by:

\begin{equation}
\hat{H}=\sum_{n}2J{\bf s}_{n}{\bf s}_{n+1}-2Ds_{n}^{z}s_{n+1}^{z}-g\mu
_{B}Hs_{n}^{y}  \label{1}
\end{equation}
with $J\simeq 6.8$ K and $D\simeq 0.16$ K and $s=5/2$. Each spin can be
described as a vector, the orientation of which is defined by two angles, $%
\theta $ and $\phi $. In configuration A, the spins are assumed to remain
within the XY plane. Accordingly, $\theta $ remains fixed to the value $\pi
/2$, and $\phi $ is defined by the angle between the spin and the external
field $H$. In the Ising-like phase of TMMC, the spin vectors is simply
written as $s_{n}^{y}=(-1)^{n}\cos (\phi $). Within a classical description
(and in the continuum limit), and after the variable change $\psi =\pi
-2\phi $ is made, hamiltonian (1) is changed into

\begin{eqnarray}
\hat{H} &=&Js^{2}/2\int_{-\infty }^{+\infty }dz/2  \nonumber \\
&&\left[ 1/C^{2}(\partial \psi /\partial t)^{2}+(\partial \psi /\partial
z)^{2}+m^{2}(1-\cos (\psi ))\right]
\end{eqnarray}
which is that of the sine-Gordon model. Here, $C=4Js[1-D/(2J)]^{1/2}$ is the
soliton velocity and $m=g\mu _{B}H/4Js$ defines the soliton mass. The
elementary excitations of this model are typical non-linear excitations
(NLE): the solitons (and antisolitons) and the breathers. A soliton in TMMC
(or an antisoliton) can be viewed as a domain wall (extending over several
lattice spacings), with a well-defined shape as represented in Fig. 2. The
breather modes are soliton-antisoliton bound-states. As a general property,
when NLE move along the chains, they maintain an exact balance between two
energy contributions, the ``potential'' and the ``internal'' energies
defined by the two last terms of (2), respectively. This explains the
remarkable ``integrity'' of the NLE excitations \cite{Scott}.

\begin{figure}[tbp]
\centerline{\epsfxsize=80 mm \epsfbox{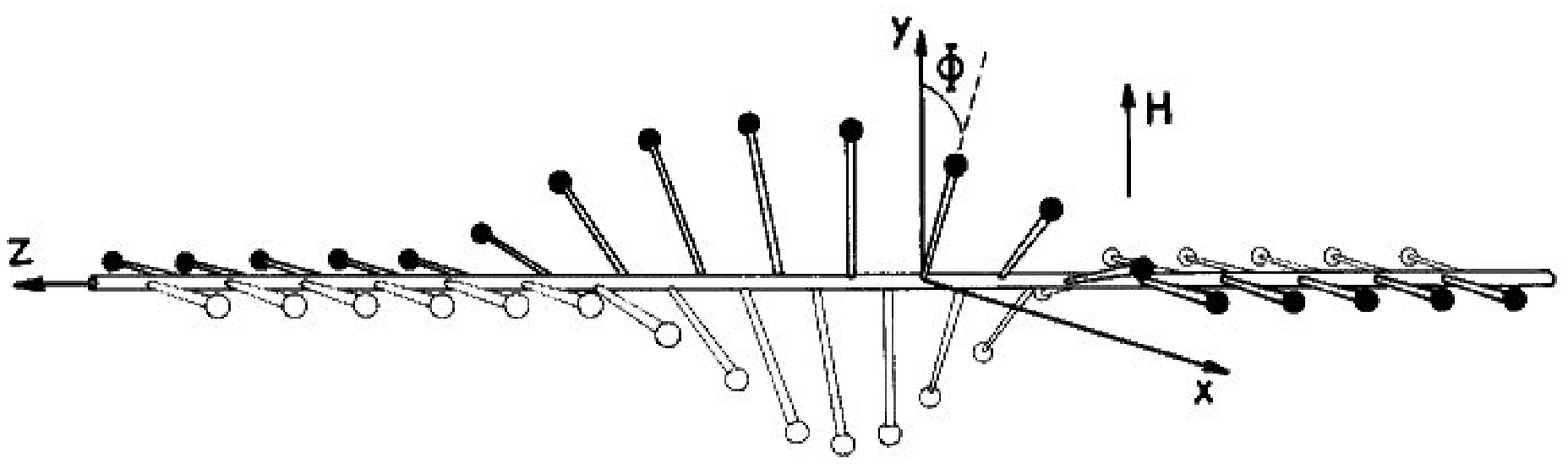}}\vspace{+0.0 cm}
\caption{Schematic representation of a (broad) soliton in TMMC.}
\end{figure}

As can be seen in Fig. 2, inside a soliton, the spins undergo a $\pi $
rotation (this provides an alternative and a more general definition of
solitons in AF chains). By looking at the fluctuations induced by the
soliton motion \cite{Boucher80,TMMC}, an accurate description of the soliton
regimes in TMMC has been obtained. Concerning the breather excitations, the
situation is experimentally more complicated. In spin systems, a
discretization of the breather spectrum is to be expected, which, then,
should result in a set of discrete excitation branches, hereafter denoted B$%
_{1}$ ... B$_{n}$. In this quantization process, the lowest breather B$_{1}$
coincides with the usual magnon excitation \cite{Maki}. If we consider only
the A soliton regime, the lowest breather gap is that of the lowest magnon
branch. The experimental gap value is in agreement with the value predicted
by (2): $E_{1}^{A}=g\mu _{B}H$. In TMMC, however, there exist two
competitive anisotropies as mentioned above. Accordingly, there is another
magnon branch with another gap: $E_{1}^{B}=4S\sqrt{D/J}$ ($\simeq 0.8$ meV).
This higher gap is that of the lowest breather branch associated with the B
soliton regime (see the insets in Fig. 1). In the quantization process, the
breathers must be considered as a whole. As a result, additional peaks can
be expected in the breather spectrum. This agrees with the observation of
``unexpected'' peaks at energies $E\simeq E_{1}^{B}\pm E_{1}^{A}$\cite
{Boucher90}.

\begin{figure}[tbp]
\centerline{\epsfxsize=80 mm \epsfbox{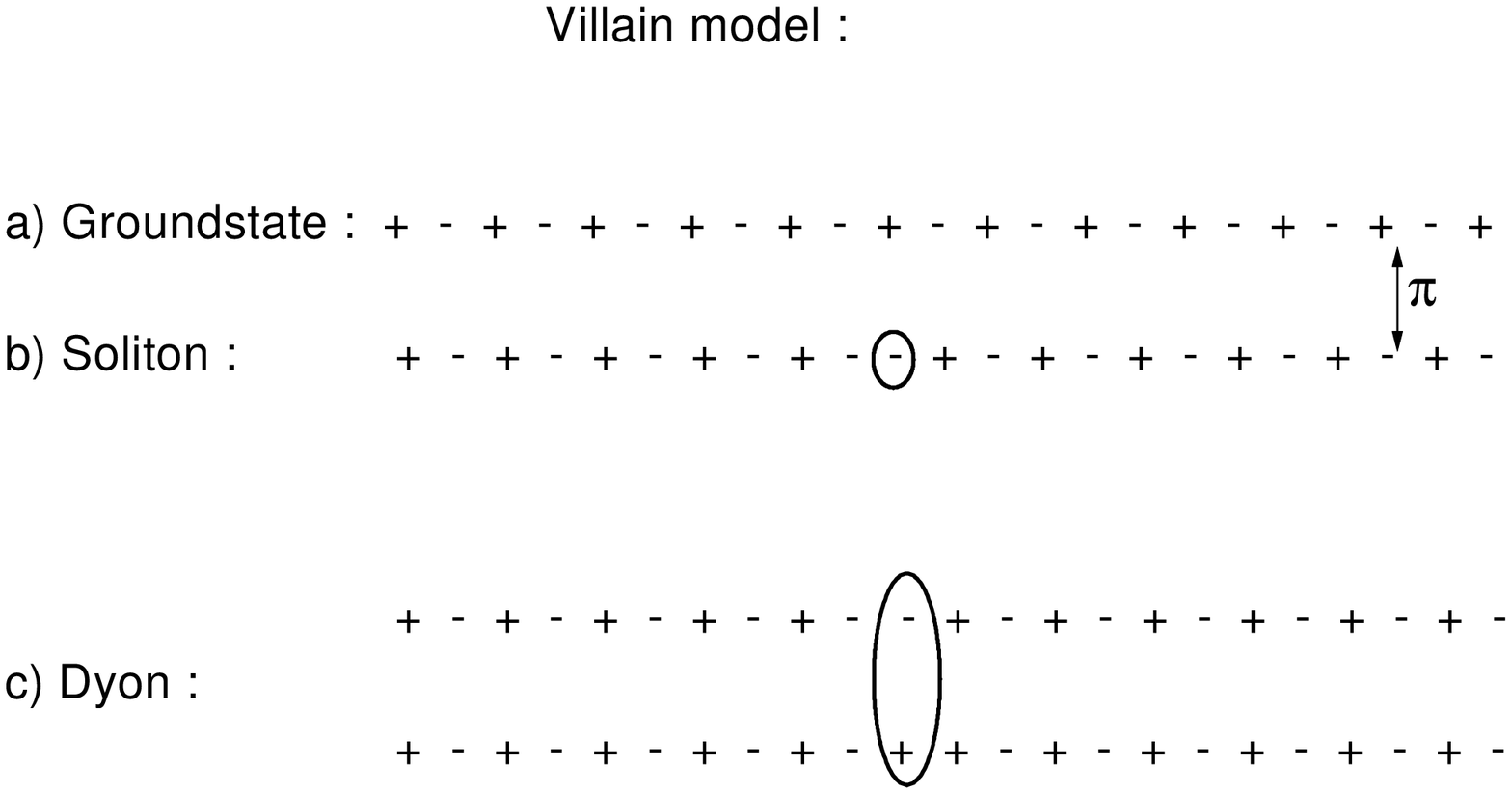}}\vspace{-5.5 cm}
\caption{Groundstate and elementary excitations in the Villain
model (hamiltonian (3)).}
\end{figure}

\section{The Co compounds: CsCoCl$_{3}$ and CsCoBr$_{3}$:}

An alternative soliton (or kink) model for AF chains has been proposed by J.
Villain, in 1975 \cite{Villain}. It applies to the following quantum ($s=1/2$%
) Ising-like Hamiltonian \cite{Devreux}:

\begin{equation}
\hat{H}=2J\sum_{n}s_{n}^{z}s_{n+1}^{z}+\varepsilon
(s_{n}^{x}s_{n+1}^{x}+s_{n}^{y}s_{n+1}^{y})  \label{3}
\end{equation}
with $\varepsilon <<1$. For this simple model, the groundstate is the
N\'{e}el state as represented in Fig. 3a.

The first excited state is displayed in Fig. 3b: it contains a narrow domain
wall (extending over only one lattice spacing) and, as expected, it is
associated with a $\pi $ rotation of the sublattices. For this model, the
dispersion of the soliton branch is shown in Fig. 4a.

\begin{figure}[tbp]
\centerline{\epsfxsize=70 mm \epsfbox{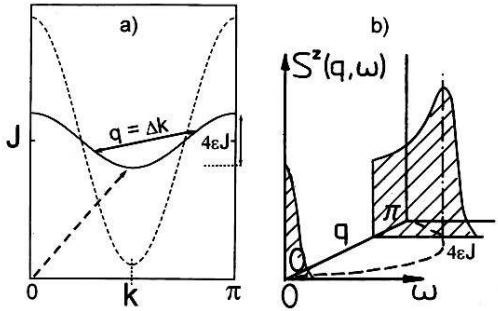}}\vspace{0 cm}
\caption{a) Dispersions of the soliton excitations in the Villain
model: the solid line is for a small $\varepsilon $ value; the
dashed line, for a large $\varepsilon $ ($\rightarrow 1$). The
arrows illustrate experimental transitions (see text). b) The
fluctuation spectrum of the (individual) soliton mode (in zero
field).}
\end{figure}

In the same figure, the arrows show the transitions to be realized in
experiments. The transitions induced from the groundstate (the dash arrow)
are forbidden. They would require that an infinite number of spins be
simultaneously flipped (in order to realize the $\pi $ rotation of the
sublattices). This establishes a pertinent result for Ising-like AF chains:
{\it the dispersion of a soliton branch cannot be determined experimentally}%
. Only transitions inside the soliton branch (shown by the full arrow) are
possible: they describe the fluctuations of (individual) solitons when they
move along the chains. The corresponding spectrum is drawn in Fig. 4b: the
fluctuations are seen to develop at very low energy (around zero energy). A
peak (i.e., a single ``maximum'') in the fluctuation spectrum is expected at
$E_{max}(q)=4\varepsilon J\sin (q)$ \cite{TMMC,Sasaki}.

\begin{figure}[tbp]
\centerline{\epsfxsize=80 mm \epsfbox{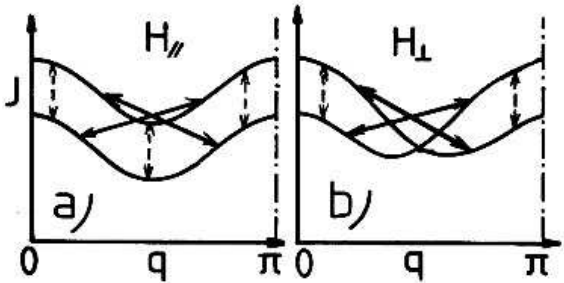}}\vspace{0 cm}
\caption{Zeeman splitting of the soliton dispersion in the Villain
model for a field parallel (a) and transverse (b) to the Ising
direction.}
\end{figure}

Such a kink in quantum spin chains is, in fact, a degenerate state. As shown
in Fig. 3c, it can be defined in two ways, which both agree with the $\pi $
rotation defined above. They differ, however, by the spin value ($\pm 1/2$)
inside the domain wall. A kink, i.e., a soliton, is a doublet state,
entirely defined by the additional quantum spin number $S=1/2$. Accordingly,
in a field $H$, a Zeeman splitting occurs in the dispersion, as shown in
Figs. 5a and b. This splitting is different for H applied parallel ($%
H_{\parallel }$) and perpendicular ($H_{\perp }$) to the Ising axis Z. The
allowed transitions are now induced between the two split branches (the full
arrows in Figs. 5). Finally, this splitting results in a ``{\it doubling}''
of the observable soliton modes with two maxima, as shown in Figs. 6a and b
\cite{Boucher87}.

\begin{figure}[tbp]
\centerline{\epsfxsize=80 mm \epsfbox{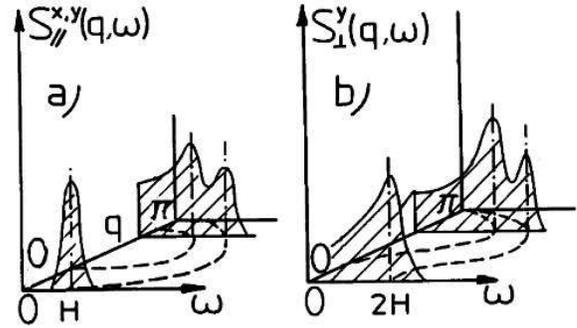}}\vspace{0 cm}
\caption{Doubling of the soliton modes in the Villain model, for H
parallel (a) and perpendicular (b) to the Ising direction.}
\end{figure}

The Co compounds, CsCoCl$_{3}$ and CsCoBr$_{3}$, are good examples of the
Villain model, with $J\simeq 75$ K and $\varepsilon \simeq 0.12$. The
fluctuations associated with the solitons have been clearly identified. In
particular, with CsCoBr$_{3}$, the low-energy fluctuations characterizing
the (individual) soliton modes have been observed by neutron inelastic
scattering (NIS) \cite{Yoshizawaka}. Evidence for a ``maximum'' in the
fluctuation spectrum requires a high instrumental resolution. Measurements
have been performed at the Institut Laue-Langevin (ILL, Grenoble, France)
using the high-flux high-resolution spectrometer IN14 \cite{Boucher90b}. An
example of the soliton mode observed in zero field for the wavevector
transfer $\Delta k=q=\pi /2$ is displayed in Fig. 7a. A single ``maximum''
is clearly detected. Its position agrees well with the expectation (the full
line is a theoretical curve which takes into account the instrumental
resolution, $\delta E\simeq 1$ meV). In Fig. 7b, the same energy scan
performed in presence of a magnetic field ($H\simeq 9.82$ T) is reported.
Two maxima are now visible: this result (the full line is the theory)
establishes the predicted doubling of the soliton modes.

The neutron data reported in Fig. 7 correspond to magnetic transitions
associated with the particular momentum transfer $\Delta k=q=\pi /2$ (the
full arrows in Figs. 4a and 5). Transitions with $\Delta k=q=0$ (the
vertical dash arrows in Fig. 5) can also be considered. They are realized in
an ESR experiment. Such measurements have been performed on the compound
CsCoCl$_{3}$ \cite{Boucher87b}. They are referred to as Soliton Magnetic
Resonance (SMR). Examples of SMR signals are reported in Fig. 8, for both $%
H_{\parallel }$ and $H_{\perp }$. As shown by the full lines
(theoretical curves), the SMR data agree also very well with this
concept of a $S=1/2$ soliton state.

\begin{figure}[tbp]
\centerline{\epsfxsize=100 mm \epsfbox{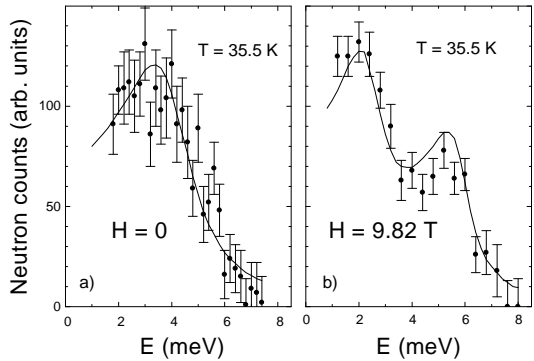}} \vspace{-8.5 cm}
\caption{NIS on CsCoBr$_{3}$: a) Soliton mode in zero field; b)
doubling of the soliton mode in a transverse field. These energy
scans correspond to the wavevector transfer $q=\pi /2$.}
\end{figure}

From these NIS and SMR measurements, an important property is established:
in quantum Ising-like chains, a soliton is a $S=1/2$ doublet excitation. In
fact, this result generalizes to any larger spin values, i.e., $s>1/2$. This
has been established, first, by Haldane in a semi-classical approach \cite
{Haldane}: a soliton in an AF chain contains internal degrees of freedom.
After quantization, they result in a discretization of the soliton states.
At this point, a drastic difference is established between integer and
half-integer spin chains: in the former case, the lowest soliton state is a $%
S=0$ excitation (this is the basis of the ``Haldane conjecture''), while in
the later case, it is a $S=1/2$ excitation (in full agreement with the
results on the Co Compounds).

\begin{figure}[tbp]
\centerline{\epsfxsize=100 mm \epsfbox{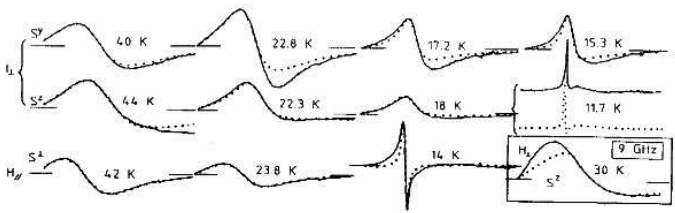}} \vspace{0 cm}
\caption{SMR signals observed in CsCoCl$_{3}$, for H parallel and
H perpendicular to the Ising direction.}
\end{figure}

Hence, in half-integer spin systems, because of these additional
degrees of freedom, the domain walls represented in Figs. 2 and
3b, are not only solitons: they are dyons, according to Affleck
\cite{Affleck86}. The dyon concept is used in particle physics: it
describes a particle with an additional ``charge'' (cf: the
magnetic monopoles). In the present NLE context, the $S=1/2$
solitons observed in the Co compounds are dyon analogs in quantum
AF chains \cite{Affleck86}.

\section{Quantum isotropic spin chains :}

Up to now, we have considered Ising-like chains, i.e., anisotropic
systems, where the NLE are rigorously defined. Within the same
context, we may try to reach a physical picture for isotropic
chains. We follow the discussion presented by Haldane for
classical spin systems \cite{Haldane}, but we apply it to the
quantum case of the Villain model. Let us consider the change
occurring in the soliton dispersion while the anisotropy is slowly
removed, i.e., when $\varepsilon \rightarrow 1$ in Eq. 3. As shown
by the dash line in Fig. 4a, the amplitude of the dispersion
increases, and at $\varepsilon =1 $, i.e., the isotropic point,
the gap of the soliton branch would close (at the momentum value
$k=\pi /2$). A phase transition occurs \cite{Singh}, which can
also be viewed as an isolant-metal transition \cite{Schulz}. The
dyon states (i.e., the pairs of $+1/2$ and $-1/2$ solitons) are
seen to fill in the groundstate of the new phase, which is that of
the quantum isotropic Heisenberg hamiltonian

\begin{equation}
\hat{H}=\sum_{n}2J{\bf s}_{n}{\bf s}_{n+1}  \label{4}
\end{equation}
For this model, another concept, the spinon concept, is commonly used \cite
{spinon}. A spinon is defined as an ``entity'' associated with a $\pm 1/2$
spin value and the groundstate of this isotropic hamiltonian is made of
pairs of spinons. Such pairs of spinons compare well with the $\pm 1/2$ dyon
states defined above.

The description of the properties of quantum Heisenberg chains started with
the Bethe ansatz, in 1931 \cite{Bethe}. In the 60's, based on the
Jordan-Wigner transformation, many studies were referring to a model of
strongly-interacting fermions \cite{Bulaevskii}. In the 70's, more
sophisticated analyzes were proposed \cite{Peschel}, which yields the modern
field-theoretic derivations, which in the recent years, have renewed and
enlarged our understandings of so many quantum spin systems. Cu Benzoate is
such a recent example. For this compound, one is led to refer also to the
sine-Gordon model \cite{Affleck99}.

In anisotropic systems, the spin direction is easily defined (for instance,
by the angle $\phi $ in Fig. 2), and/or with respect to the groundstate. The
pertinent variable in the sine-Gordon hamiltonian of Eq. 2 remains this
angle $\phi $. In isotropic systems, however, there is no definite direction
for the spins. Alternative descriptions have been proposed \cite{Peschel}.
Usually, they starts with the Jordan-Wigner transformation. This leads to a
fermionic representation of the hamiltonian with, in reciprocal space, the
fermions operators $\alpha _{q}$ ($\alpha _{q}^{+}$) and $\beta _{q}$ ($%
\beta _{q}^{+}$). It is then recognized that the significant physical
quantities are not contained in the fermion operators but rather in the
fermion densities: $\rho _{q}^{+}\simeq \sum_{k}\alpha _{k+q}^{+}\alpha _{k}$
and $\rho _{q}^{-}\simeq \sum_{k}\beta _{k+q}^{+}\beta _{k}$ (this is the
so-called ``bozonization''). Another transformation - this is the basis of
the field theoretic approach - brings back the equations to the real space
where a field ``$\widetilde{\psi }$'' is defined. Applying this procedure to
the isotropic hamiltonian (4) leads to

\begin{equation}
\hat{H}=J/8\int_{-\infty }^{+\infty }dz1/2\left[ 1/\widetilde{C}%
^{2}(\partial \widetilde{\psi }/\partial t)^{2}+(\partial \widetilde{\psi }%
/\partial z)^{2}\right]  \label{5}
\end{equation}
with $\widetilde{C}=\pi J$. In this expression, the field $\widetilde{\psi }$
does not compare to the angle $\psi $ introduced in Eq. 2: $\widetilde{\psi }
$ is related to the fermion densities $\rho ^{+}$ and $\rho ^{-}$ defined
above. Eq. 5 is not the sine-Gordon Hamiltonian. The last term which, in Eq.
2, introduces the non-linearity is missing. Accordingly, the excitations
corresponding to Hamiltonian (5) follow {\it linear} dispersions (i.e., {\it %
no energy gap}). They describe the lowest part of the spinon continuum,
which characterizes the excitation spectrum of (4). As it is known, the
application of a field H on hamiltonian (4) develops a ``{\it dynamical
incommensurability}'': the wavevectors where the critical zero-energy
fluctuations take place are shifted by $\delta q_{inc}=2\pi \sigma $, where $%
\sigma $, which is the magnetization per spin, is an increasing function of $%
H$. The dispersions, however, remain linear at low energy and no
energy gap is induced by a field.

\section{Cu Benzoate:}

In the 70's, CuBenz was considered as a good example of quantum isotropic AF
chains \cite{Okuda}. Recently, however, it has been observed that, in a
field, an energy gap opens in the energy spectrum \cite{Dender}, in
contradiction with the expectation for the isotropic model. According to
Affleck and Oshikawa \cite{Affleck99}, this behavior is due to the presence
of an additional ``staggered'' field, $h_{st}$ \cite{DM}. In a field, the
effective hamiltonian for Cu Benzoate is to be written:

\begin{equation}
\hat{H}=\sum_{n}2J{\bf s}_{n}{\bf s}_{n+1}-g\mu _{B}Hs_{n}^{y}-g\mu
_{B}h_{st}s_{n}^{x}  \label{6}
\end{equation}
with $J\simeq 5.8$ K, and where $h_{st}$ is proportional to the applied
field: $h_{st}\simeq 0.09H$ (for $H$ applied parallel to the c axis) \cite
{Oshikawa97}. The field-theoretic derivation of this hamiltonian\cite
{Affleck99} transforms Eq. 6 into the sine-Gordon equation:

\begin{eqnarray}
\hat{H} &=&J/8\int_{-\infty }^{+\infty }dz/2  \nonumber \\
&&\left[ 1/\widetilde{C}^{2}(\partial \widetilde{\psi }/\partial
t)^{2}+(\partial \widetilde{\psi }/\partial z)^{2}+\widetilde{m}^{2}(1-\cos (%
\widetilde{\psi }))\right]  \label{7}
\end{eqnarray}
where $\widetilde{C}{}$ and the field $\widetilde{\psi }$ are
defined in a similar way as in Eq. 5. The soliton mass is now a
function of the staggered field: $\widetilde{m}\simeq
1.85(h_{st}/J)^{2/3}\left| \ln (h_{st}/J)\right| $
\cite{Affleck99}. The low-energy excitations of (7) are solitons
(antisolitons) and breathers.

\begin{figure}[tbp]
\centerline{\epsfxsize=92 mm \epsfbox{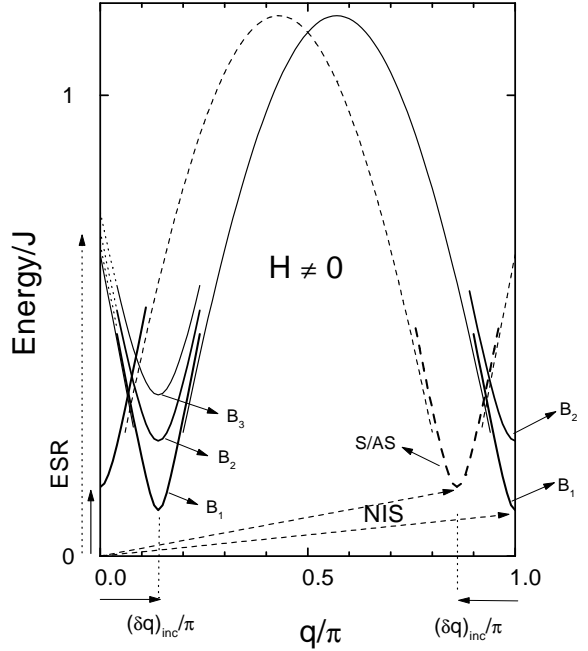}} \vspace{-3.5 cm}
\caption{A model for the dispersion of the elementary excitations
in Cu Benzoate: the soliton/antisoliton branch gives a gap at $q =
0$ and at the incommensurate wavevector $q=\pi-\delta q_{inc}$.
The breather branches give a gap at $q =\pi$ and at the
incommensurate wavevector $q = \delta q_{inc}$. The dashed arrows
corresponds to NIS [26], the full arrows to ESR [3].}
\end{figure}

In Fig. 9, a complete representation of the elementary excitations
in Cu Benzoate is proposed. The bold lines correspond to the
sine-Gordon predictions (low-energy limit). In this figure, three
breather branches are represented: $B_{1}$, $B_{2}$ and $B_{3}$.

We now review briefly the recent experiments performed in Cu
Benzoate. In Fig. 9, the transitions observed by NIS are shown by
the dash arrows (these measurements have been performed at a
single field value $H\simeq 7$ T) \cite {Dender}. The dynamical
incommensurability has been observed at $q=\pi -\delta q_{inc}$.
At this precise wavevector value, the gap of the
soliton/antisoliton branch is directly measured. At $q=\pi $, the
two first breather modes, $B_{1}$ and $B_{2}$, have been
observed\cite{Essler}. Recently, ESR measurements have also been
performed in Cu Benzoate \cite {Asano}. In fig. 9, the transitions
observed by ESR are shown by the vertical dotted and solid arrows.
The solid arrow shows that the soliton/antisoliton gap is measured
at $q=0$. The dotted arrow probes the breather modes, also at
$q=0$ \cite{Oshikawa99}. All these ESR measurements have been
performed in a wide field range, up to $H\simeq 20$ T. The
observed field dependence of the soliton gap is displayed in Fig.
10. In low field, the agreement with the sine-Gordon model (the
full line) is very good: $E_{G}=\widetilde{m}J\simeq H^{2/3}$. An
appreciable deviation, however, occurs in high fields. In Fig. 11,
the field dependence of the three first breathers, $B_{1}$,
$B_{2}$ and $B_{3}$ as probed by ESR (i.e., at q = 0) is obtained
in the full field range.

\begin{figure}[tbp]
\centerline{\epsfxsize=80 mm \epsfbox{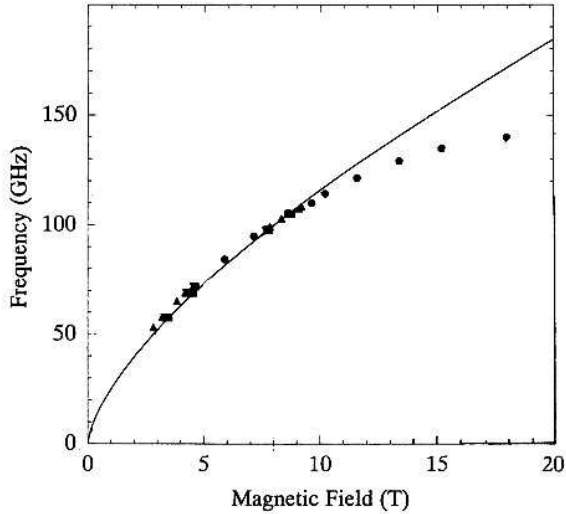}} \vspace{0.2 cm}
\caption{Field dependence of the soliton gap observed by ESR
(Voigt configuration) in Cu Benzoate.}
\end{figure}

\section{Conclusion:}

With these compounds - i.e., TMMC, the Co compounds and Cu Benzoate - the
importance of the NLE in the low-T properties of AF chains is well
exemplified. They are seen to play a crucial role in both anisotropic and
isotropic systems. They come in the description of both quantum and
classical spin chains. TMMC and Cu Benzoate are both well described by the
sine-Gordon model. Fundamental differences, however, occur between the two
systems, in particular for the experiments.

\begin{figure}[tbp]
\centerline{\epsfxsize=90 mm \epsfbox{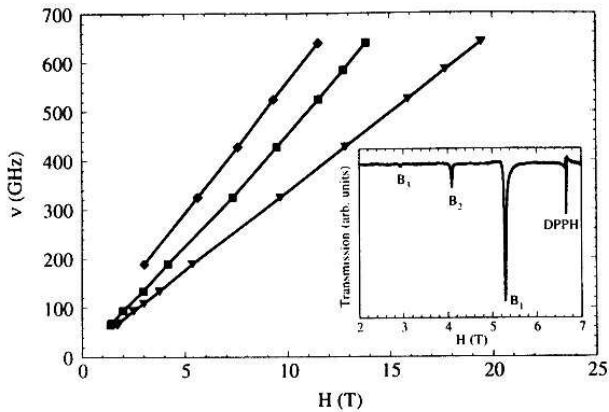}} \vspace{+0 cm}
\caption{Field dependence of the first breather modes observed by
ESR (Faraday configuration) in Cu Benzoate [3].}
\end{figure}

In Ising-like chains (TMMC and the Co compounds), transitions from
the groundstate to the soliton state are forbidden (see the dash
line in Fig. 4a). In such systems, the presence of solitons is
detected by looking at the low-energy fluctuations induced by the
soliton motion \cite{TMMC,Yoshizawaka}. In Cu Benzoate, however,
the soliton dispersion can be determined experimentally: the
transitions from the groundstate to both the soliton and the
breather branches are allowed, (see the arrows in Fig. 9). With Cu
Benzoate, the discretization of the breather spectrum in a
sine-Gordon model is now well established. \newline
\newline
Acknowledgements: One of us (JPB) would like to thank T. Ziman for
illuminating discussions.


\begin{references}
\bibitem{Scott}  A. C. Scott, F.Y.F. Chu, and D. W. McLaughlin, Proceedings
of the IEEE, {\bf 61}, 1443, (1973).

\bibitem{Boucher80}  J.P. Boucher, L.P. Regnault, J. Rossad-Mignod, J.P.
Renard, J. Bouillot and W.G. Stirling, Sol. Stat. Comm., {\bf 33}, 171
(1980).

\bibitem{Asano}  T. Asano, H. Nojiri, Y. Inagaki, J.P. Boucher, T. Sakon, Y.
Ajiro, and M. Motokawa, Phys. Rev. Lett. {\bf 84}, 5880 (2000).

\bibitem{Boucher}  For a review on solitons in TMMC and the Co compounds,
see for instance: J.P. Boucher, L.P. Regnault, and H. Benner, in
Nonlinearity in Condensed Matter, Springer Series in Solid State-Sciences,
Vol. {\bf 69}, 24 (1987).

\bibitem{Boucher79}  J.P. Boucher, L.P. Regnault, J. Rossad-Mignod and J.
Villain, J. M. M. M., {\bf 14}, 155, (1979).

\bibitem{TMMC}  In TMMC, both the fluctuations associated with the flipping
process of the sublattices, and those associated with the individual
solitons have been observed: J.P. Boucher, L.P. Regnault, R. Pynn, J.
Bouillot, and J.P. Renard, Europhys. Lett. {\bf 1}, 415 (1986).

\bibitem{Maki}  K. Maki and H. Takayama, Phys. Rev. B {\bf 22}, 5302 (1980).

\bibitem{Boucher90}  J.P. Boucher, R. Pynn, M. Remoissenet, L.P. Regnault,
Y. Endoh, and J.P. Renard, Phys. Rev. Lett., {\bf 64}, 1557 (1990).

\bibitem{Villain}  J. Villain, Physica B {\bf 79}, 1 (1975).

\bibitem{Devreux}  For the Villain model, several derivations have been
proposed. In a second quantization approach, the solitons have been shown to
be fermions: F. Devreux and J.P. Boucher, J. Phys. (France) {\bf 48}, 1663
(1987).

\bibitem{Sasaki}  A limitation of the diverging peak results from the
collisions between solitons: K. Sasaki and K. Maki, Phys. Rev. B {\bf 35},
257 (1987).

\bibitem{Boucher87}  J.P. Boucher, in Magnetic Excitations and Fluctuations
II, Proceedings in Physics {\bf 23}, p. 171, (1987), edited by: U. Bucalani,
S. W. Lovesey, M.G. Rasetti and V. Tognetti, Springer Verlag.

\bibitem{Yoshizawaka}  H. Yoshizawa, K. Hirakawa, S.K. Satija, and G.
Shirane, Phys. Rev. B {\bf 23}, 2298 (1981); S.E. Nagler, W.J.L. Buyers,
R.L. Armstrong, and B. Briat, Phys. Rev. Lett. {\bf 49}, 590 (1982); J.P.
Boucher, L.P. Regnault, J. Rossad-Mignod, J.Y. Henry, J. Bouillot and W.G.
Stirling, Phys. Rev. B {\bf 31}, 3015 (1985).

\bibitem{Boucher90b}  J.P. Boucher, L.P. Regnault, R. Currat, J.Y. Henry,
ILL Report 1990.

\bibitem{Boucher87b}  J.P. Boucher, G. Rius, and J.Y. Henry, Europhys. Lett.
{\bf 4}, 1073 (1987).

\bibitem{Haldane}  F.D.M. Haldane, Phys. Rev. Lett. {\bf 50}, 1153 (1983).

\bibitem{Affleck86}  I. Affleck, Phys. Rev. Lett. {\bf 57}, 1048 (1986).

\bibitem{Singh}  R.R.P. Singh, Phys. Rev. B {\bf 53}, 11582 (1996).

\bibitem{Schulz}  H. J. Schulz, in Proceedings of Les Houches Summer School
LXI, ed. E. Akkermans, G. Montambaux, J. Pichard, et J. Zinn-Justin
(Elsevier, Amsterdam, 1995), p.533.

\bibitem{spinon}  L.D. Fadeev and L.A. Takhtajan, Phys. Lett. {\bf 85A}, 375
(1981).

\bibitem{Bethe}  H. Bethe, Z Physik {\bf 71}, 205 (1931).

\bibitem{Bulaevskii}  For instance, L.N. Bulaevskii, Sov. Phys. JETP {\bf 16}%
, 685 (1963).

\bibitem{Peschel}  For instance, A. Luther and I. Peschel, Phys. Rev. B {\bf %
12}, 3908 (1975); J. Solyom, in Lectures notes in Physics 96, Quasi
one-Dimensional II, p. 20, 1978; Springer-Verlag.

\bibitem{Affleck99}  I. Affleck and M. Oshikawa, Phys. Rev. B {\bf 60}, 1038
(1999).

\bibitem{Okuda}  K. Okuda, H. Hata, and M. Date, J. Phys. Soc. Jpn. {\bf 33}%
, 1574 (1972).

\bibitem{Dender}  D.C. Dender, P.R. Hammar, D.H. Reich, C. Broholm and G.
Aeppli, Phys. Rev. Lett., {\bf 79}, 1750 (1997).

\bibitem{DM}  This staggered field results from two contributions: one is
due to the g factor, which undergoes a small alternation along the chain,
the other to the presence of small Dzyaloshinski-Moryia interactions (see
[24]).

\bibitem{Oshikawa97}  M. Oshikawa and I. Affleck, Phys. Rev. Lett. {\bf 79},
2883 (1997).

\bibitem{Essler}  F.H.L. Essler and A. M. Tsvelik, Phys. Rev. B {\bf 57},
10592 (1998).

\bibitem{Oshikawa99}  M. Oshikawa and I. Affleck, Phys. Rev. Lett. {\bf 82},
5136 (1999).
\end{references}
\end{document}